\documentclass[apj]{emulateapj} 

\usepackage{epsfig}
\usepackage{amsmath}
\usepackage{amssymb}
\usepackage{natbib}
\usepackage{threeparttable} 
\usepackage{booktabs}

\usepackage{epstopdf}

\newcommand{\chan}{\textit{Chandra}}

\newcommand{\Msun}{\mathrm{M}_{\odot}}
\newcommand{\lum}{\mathrm{erg~s}^{-1}}
\newcommand{\flux}{\mathrm{erg~cm}^{-2}~\mathrm{s}^{-1}}

\newcommand{\cnts}{\mathrm{counts~s}^{-1}}
\newcommand{\mdot}{\mathrm{M_{\odot}~yr}^{-1}}
\newcommand{\nh}{\mathrm{cm}^{-2}}

\newcommand{\exo}{EXO 0748--676}
\newcommand{\igr}{IGR J17480--2446}

\newcommand{\xte}{XTE J1701--462}

\newcommand{\ks}{KS~1731--260}
\newcommand{\mxb}{MXB~1659--29}
\newcommand{\rx}{XTE~J1709--267}
\newcommand{\maxisource}{MAXI~J0556--332}
\newcommand{\terdrie}{Swift~J174805.3--244637}

\def \nar {NewAR}

\slugcomment{Received 2013 June 10; accepted 2013 August 7}

\shorttitle{Crust cooling of \igr}
\shortauthors{Degenaar et al.}

\begin{document}

\title{Continued Neutron Star Crust Cooling of the 11 Hz X-Ray Pulsar in Terzan 5:\\ A Challenge to Heating and Cooling Models?}

\author{N.~Degenaar$^{1,}$\altaffilmark{9}, R.~Wijnands$^{2}$, E.F.~Brown$^{3}$, D.~Altamirano$^2$, E.M.~Cackett$^4$, J.~Fridriksson$^2$, J.~Homan$^5$, C.O.~Heinke$^6$, J.M.~Miller$^{1}$, D.~Pooley$^{7,8}$, G.R.~Sivakoff$^6$
}
\affil{$^1$Department of Astronomy, University of Michigan, 500 Church Street, Ann Arbor, MI 48109, USA; degenaar@umich.edu\\
$^2$Astronomical Institute Anton Pannekoek, University of Amsterdam, Postbus 94249, 1090 GE Amsterdam, The Netherlands\\
$^3$Department of Physics and Astronomy, Michigan State University, East Lansing, MI 48824, USA\\
$^4$Department of Physics and Astronomy, Wayne State University, 666 W. Hancock St, Detroit, MI 48201, USA\\
$^5$Massachusetts Institute of Technology, Kavli Institute for Astrophysics and Space Research, Cambridge, MA 02139, USA\\
$^6$Department of Physics, University of Alberta, 4-183 CCIS, Edmonton, AB T6G 2E1, Canada\\
$^7$Department of Physics, Sam Houston State University, Huntsville, TX, USA\\
$^8$Eureka Scientific, Inc., 2452 Delmer Street, Suite 100, Oakland, CA 94602, USA
}
\altaffiltext{9}{Hubble Fellow}

\begin{abstract}
The transient neutron star low-mass X-ray binary and 11 Hz X-ray pulsar \igr\ in the globular cluster Terzan 5 exhibited an 11-week accretion outburst in 2010. \chan\ observations performed within five months after the end of the outburst revealed evidence that the crust of the neutron star became substantially heated during the accretion episode and was subsequently cooling in quiescence. This provides the rare opportunity to probe the structure and composition of the crust. Here, we report on new \chan\ observations of Terzan 5 that extend the monitoring to $\simeq$2.2~yr into quiescence. We find that the thermal flux and neutron star temperature have continued to decrease, but remain significantly above the values that were measured before the 2010 accretion phase. This suggests that the crust has not thermally relaxed yet, and may continue to cool. Such behavior is difficult to explain within our current understanding of heating and cooling of transiently accreting neutron stars. Alternatively, the quiescent emission may have settled at a higher observed equilibrium level (for the same interior temperature), in which case the neutron star crust may have fully cooled.
\end{abstract}



\section{Introduction}
Studying the thermal evolution of neutron stars is a promising avenue to gain insight into their structure and composition. These compact stellar remnants are born hot in supernova explosions, but quickly cool as their thermal energy is drained via neutrino emission from their dense interior and thermal photons radiated from their surface. 
When residing in low-mass X-ray binaries (LMXBs), neutron stars accrete matter from a late-type companion star that overflows its Roche lobe. The accretion of matter can re-heat the neutron star and drastically impact its thermal evolution.

Accretion causes the original crust of a neutron star \citep[built of cold, catalyzed matter;][]{shapiro1986} to become replaced by one formed by the processed material. An accreted crust is out of nuclear equilibrium and represents a reservoir of energy as it provides a site for non-equilibrium processes that generate heat \citep[e.g.,][]{sato1979,haensel1990b,haensel1990a,steiner2012}. 
Compression of the crust by ongoing accretion induces a chain of nuclear reactions that produce heat at a rate that is proportional to the mass-accretion rate. In the outer crustal layers, electron captures generate on the order of $\simeq0.01$~MeV per accreted nucleon \citep[e.g.,][]{gupta07,estrade2011}. However, most heat is produced in pycnonuclear fusion reactions that occur deep within the crust and release $\simeq1.5$~MeV nucleon$^{-1}$ \citep[e.g.,][]{haensel1990a,yakovlev2006,horowitz2008}. The structure and composition of the crust play an important role in the heat generation.

In transient LMXBs, a neutron star is typically accreting actively only for a few months at a time. Such accretion outbursts are interleaved by quiescent episodes, generally lasting for years or decades, during which accretion onto the neutron star is strongly reduced or completely halted. During these quiescent intervals, the accretion-heated crust cools as the gained energy is thermally conducted toward the stellar core and surface \citep[e.g.,][]{ushomirsky2001,rutledge2002}. As the crust thermally relaxes, it eventually settles at a stable level that is determined by the core temperature, which evolves on a much longer time scale \citep[$\simeq10^{4}$~yr; e.g.,][]{colpi2001}. The cooling rate is sensitive to the heat capacity and thermal transport properties of the crust, and hence to its structure and composition.

Multi-epoch observations of four neutron star LMXBs following long ($>1$~yr) accretion outbursts have revealed a steady decrease in thermal X-ray emission on a time scale of years \citep[e.g.,][]{wijnands2002,wijnands2004,cackett2008,cackett2010,degenaar2010_exo2,diaztrigo2011,fridriksson2011}. These observations can successfully be modeled as cooling of an accretion-heated neutron star crust and have provided valuable insights into the properties of these layers \citep[][]{rutledge2002,shternin07,brown08,page2012}. These four so-called quasi-persistent LMXBs served as prime targets because their prolonged accretion phases were expected to severely heat the crust so that the subsequent cooling would become detectable. However, shorter outbursts can potentially also cause significant crustal heating \citep[][]{brown1998}.

\igr\ is a neutron star LMXB that contains an 11 Hz X-ray pulsar and a $\simeq$0.8$~\Msun$ companion star \citep[][]{strohmayer2010_2,testa2012}. The source is located in the dense core of the globular cluster Terzan 5, which lies at an estimated distance of $D\simeq5.5$~kpc \citep[][]{ortolani2007}. \igr\ was identified as a transient X-ray source when it entered an accretion outburst in 2010 October \citep[][]{bordas2010,pooley2010}. It remained active for $\simeq$11 weeks at an estimated average bolometric luminosity of $L_X \simeq 6\times 10^{37}~\lum$ \citep[][]{linares2012_ter5_2}, before it returned to quiescence in 2010 December \citep[][]{degenaar2011_terzan5_2}.

\chan\ observations obtained in 2011 February, $\simeq$50~days after the end of the outburst, revealed that the quiescent X-ray emission of \igr\ was elevated compared to the level measured from archival observations taken in 2003 and 2009 \citep[][]{degenaar2011_terzan5_2,degenaar2011_terzan5}. A new observation obtained $\simeq$75~days later, in 2011 April, revealed that the thermal emission had decreased but still remained above the 2003/2009 level. By invoking the presence of a strong, additional heat source in the outer crustal layers, it was proposed that the crust became significantly heated during the accretion phase and was subsequently cooling in quiescence \citep[][]{degenaar2011_terzan5_3}. In this work we present new X-ray observations of Terzan 5 that support this hypothesis.


\section{Observations, Data Analysis and Results}
Between 2011 September and 2013 February, six new \chan/ACIS observations were performed of Terzan 5 at times when no bright X-ray transients were active (Table~\ref{tab:obs}). These can be used to further study the thermal evolution of the 11 Hz X-ray pulsar. All observations were performed in the ``faint'' timed data mode with the globular cluster positioned on the S3 chip. A 1/4 sub-array was used for the observations with IDs 14475--77, whereas all others were carried out in full frame mode. All observations were free from background flares. We reduced and analyzed the data using the \textsc{ciao} tools version 4.5 and \textsc{caldb} version 4.5.5. 

We extracted source events from \igr\ by using a circular region with a radius of $1''$. As a background reference we used a source-free circular region with a radius of $40''$ that was positioned $\simeq 1.4'$ west of the cluster core. Count rates were extracted using \textsc{dmextract}, whereas spectra and the associated response files were created using \textsc{specextract}. We grouped the spectral data to contain a minimum of 15 photons per bin using \textsc{grppha}, and performed spectral fits in the 0.5--10 keV range using \textsc{XSpec} version 12.7 \citep[][]{xspec}. All uncertainties quoted in the text and presented in plots and tables are at the 1$\sigma$ level of confidence.

The 2011 September observation was split into two exposures that were taken within three days (Obs IDs 13705 and 14339; Table~\ref{tab:obs}). \igr\ is detected at similar count rates in both observations and we did not find any significant spectral differences when analyzing the two data sets separately. We therefore summed the two spectra and weighted response files using the task \textsc{combine$\_$spectra} to improve the statistics. Likewise, we combined the three exposures that were taken in 2013 February within an interval of 19 days (Obs IDs 14477, 14625, and 15615; Table~\ref{tab:obs}).

\begin{table}
\begin{center}
\caption{New \chan/ACIS-S Observations of Terzan 5\label{tab:obs}}
\begin{tabular*}{0.49\textwidth}{@{\extracolsep{\fill}}cccc}
\hline\hline
Obs ID & Date & Exposure Time  & Count Rate \\
& & (ks) &  ($10^{-3}~\cnts$) \\
\hline
13705 & 2011 Sep 5 & 13.9 & $3.41\pm0.50$ \\
14339 & 2011 Sep 8 & 34.1 & $3.06\pm0.30$ \\
13706 & 2012 May 13 & $46.5$ & $2.60\pm0.24$ \\
14475 & 2012 Sep 17/18 & $30.5$ & $2.92\pm0.31$ \\
14476 & 2012 Oct 28 & $28.6$ & $2.42\pm0.29$ \\
14477 & 2013 Feb 5 & $28.6$ & $1.76\pm0.25$ \\
14625 & 2013 Feb 22 & $49.2$ & $2.07\pm0.21$ \\
15615 & 2013 Feb 24 & $84.2$ & $1.90\pm0.15$ \\
\hline
\end{tabular*}
\tablecomments{The count rates of \igr\ are given for the 0.3--10 keV energy range. Quoted uncertainties are at the 1$\sigma$ level of confidence.}
\end{center}
\end{table}


\subsection{Spectral Analysis}
We fit all data sets simultaneously to study the thermal evolution of the neutron star. To ensure a homogenous analysis, we include the \chan\ observations performed in 2003 and 2009 (i.e., before the 2010 accretion outburst; Obs IDs 3798 and 10059), as well as those obtained in 2011 February and April (Obs IDs 13225 and 13252). For details on those observations, we refer to \citet{degenaar2011_terzan5_2,degenaar2011_terzan5} and \citet{degenaar2011_terzan5_3}. 

Previous studies showed that the quiescent spectra of \igr\ were fitted well with a thermal emission model \citep[][]{degenaar2011_terzan5_2,degenaar2011_terzan5,degenaar2011_terzan5_3}. We use the neutron star atmosphere model \textsc{nsatmos} of \citet{heinke2006}, for which we fix the mass and radius of the neutron star at $M=1.4~\Msun$ and $R=10$~km, the source distance at $D=5.5$~kpc, and the normalization at unity (which implies that the entire neutron star is radiating).  As such, the only free fit parameter for this model is the neutron star effective temperature $kT$. 
Since it is common in the literature to quote the temperature as seen by a distant observer, we convert the fitted temperatures to $kT^{\infty}= kT/(1+z)$, where $1+z = (1-R_{\mathrm{s}}/R)^{-1/2} = 1.31$ is the gravitational redshift factor for our choice of $M$ and $R$ (with $R_{\mathrm{s}}=2GM/c^2$ being the Schwarzschild radius, $G$ the gravitational constant, and $c$ the speed of light). 

In all spectral fits we account for interstellar absorption by including the \textsc{tbabs} model \citep[][]{wilms2000} with the \textsc{vern} cross sections \citep[][]{verner1996} and \textsc{wilm} abundances \citep[][]{wilms2000}. We tie the hydrogen column density between the different observations, i.e., this parameter is assumed to be constant at all epochs \citep[for a justification, see][]{miller2009}. The \textsc{nsatmos} model fits were extrapolated to the 0.01--100 keV range to obtain an estimate of the (unabsorbed) thermal bolometric flux. The results of our spectral analysis are summarized in Table~\ref{tab:spec}.

Fitting all quiescent spectral data simultaneously yields a good fit with a reduced chi-squared value of $\chi_{\nu}^2=0.89$ for 61 degrees of freedom (dof) and a $p$-value of $P_{\chi}$=0.72.\footnote[10]{The $p$-value associated with the test statistic represents the probability that deviations between the model and the data are due to chance alone. Generally, the model is rejected when $P_{\chi}\lesssim0.05$.} The obtained hydrogen column density, $N_H = (1.98\pm0.07)\times 10^{22}~\nh$, is consistent with the average value found for the 16 brightest X-ray point sources in the cluster \citep[][]{heinke2006_terzan5}, and the values obtained in previous quiescent studies of \igr\ \citep[][]{degenaar2011_terzan5_2,degenaar2011_terzan5,degenaar2011_terzan5_3}. 

The obtained temperatures show a gradual decrease from $kT^{\infty}\simeq100$ to $\simeq83$~eV over the $\simeq2.2$~yr time span covered by the observations. These values are higher than those measured from pre-outburst data obtained in 2003/2009 ($kT^{\infty}\simeq74$~eV). The inferred 0.5--10 keV thermal luminosity decreases from $L_{\rm X}\simeq1.5\times 10^{33}$ to $\simeq6.3\times 10^{32}~\lum$, but remains above the pre-outburst level of $L_{\rm X}\simeq3.6\times 10^{32}~\lum$. The estimated bolometric flux is on average a factor $\simeq1.6$ higher than the flux measured in the 0.5--10 keV band (Table~\ref{tab:spec}). The temperatures and fluxes obtained for the 2003/2009 and 2011 February/April data are consistent with the values reported in previous work \citep[][]{degenaar2011_terzan5_2,degenaar2011_terzan5,degenaar2011_terzan5_3}.

We note that the \textsc{nsatmos} model assumes that the magnetic field has a negligible effect on the emerging spectrum, which is justified for $B\lesssim10^9$~G \citep[][]{heinke2006}. However, this may not be valid for \igr, as it has an estimated magnetic field of $B\simeq10^9-10^{10}$~G \citep[][]{cavecchi2011,miller2011,papitto2010}. Magnetized neutron star atmosphere models, however, only allow for much higher field strengths of $B\geq10^{12}$~G. We briefly explored one such model \citep[\textsc{nsa};][]{zavlin1996}, adopting $B=10^{12}$~G. This did not yield an acceptable fit to the combined data set ($\chi_{\nu}^2=1.71$ for 88 dof, $P_{\chi}=3.6\times10^{-5}$). However, since we compare relative fluxes and temperatures, the observed decrease should be robust and not caused by any model uncertainties (nor by systematic uncertainties such as the source distance).

 \begin{figure}
 \begin{center}
	\includegraphics[width=8.8cm]{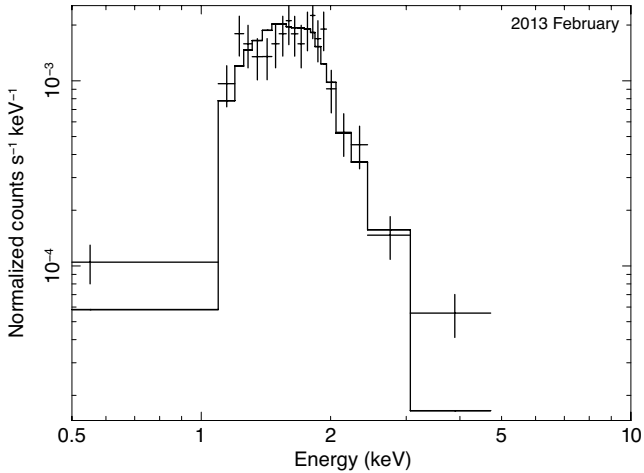}
    \end{center}
    \caption[]{Combined X-ray spectrum of the three observations performed in 2013 February (Obs IDs 14477, 14625, and 15615). The solid line represents a fit using an absorbed neutron star atmosphere model.  
        }
 \label{fig:spec}
\end{figure}

\begin{table*}
\begin{center}
\caption{Results from Analysis of the Spectral Data\label{tab:spec}}
\begin{tabular*}{1.0\textwidth}{@{\extracolsep{\fill}}cccccccc}
\hline\hline
Epoch & MJD & $kT^{\infty}$  & $F_{\mathrm{X}}$ & $F_{\mathrm{bol}}$ & $L_{\mathrm{X}}$ & $L_{\mathrm{bol}}$ \\
& & (eV) & \multicolumn{2}{c}{($10^{-13}~\flux)$} & \multicolumn{2}{c}{($10^{32}~\lum)$} \\
\hline
2003/2009 & 52833.5/55027.5 & $73.6 \pm 1.6$ & $1.00 \pm 0.12$ & $1.78 \pm 0.20$ & $3.62 \pm 0.43$ & $6.44 \pm 0.72$ \\
2011 Feb & 55609 & $99.7 \pm 1.6$ & $4.19 \pm 0.33$ & $6.02 \pm 0.41$ & $15.2 \pm 1.2$ & $21.8 \pm 1.5$ \\
2011 Apr & 55680.5 & $91.5 \pm 1.5$ & $2.81 \pm 0.23$ & $4.26 \pm 0.29$ & $10.2 \pm 0.8$ & $15.4 \pm 1.0$ \\
2011 Sep & 55810.5 & $89.2 \pm 1.5$ & $2.50 \pm 0.19$ & $3.86 \pm 0.25$ & $9.05 \pm 0.69$ & $14.0 \pm 0.9$ \\
2012 May & 56060 & $84.8 \pm 1.5$ & $1.97 \pm 0.19$ & $3.15 \pm 0.23$ & $7.13 \pm 0.69$ & $11.4 \pm 0.8$ \\
2012 Sep  & 56187.5 & $88.5 \pm 1.9$ & $2.40 \pm 0.24$ & $3.73 \pm 0.32$ & $8.69 \pm 0.87$ & $13.5 \pm 1.2$ \\
2012 Oct  & 56228 & $84.6 \pm 2.0$ & $1.94 \pm 0.23$ & $3.11 \pm 0.32$ & $7.02 \pm 0.83$ & $11.3 \pm 1.2$ \\
2013 Feb & 56340 & $82.8 \pm 1.2$ & $1.75 \pm 0.12$ & $2.86 \pm 0.17$ & $6.33 \pm 0.43$ & $10.4 \pm 0.6$ \\
\hline
\end{tabular*}
\tablecomments{$F_{\mathrm{X}}$ and $F_{\mathrm{bol}}$ represent the unabsorbed fluxes in the 0.5--10 keV and 0.01--100 keV bands, respectively. $L_{\mathrm{X}}$ and $L_{\mathrm{bol}}$ give the corresponding luminosities for a distance of $D=5.5$~kpc. 
For the spectral fits, the neutron star mass and radius were fixed at $M=1.4~\Msun$ and $R=10$~km, and a distance of $D=5.5$~kpc was assumed. The hydrogen column density was tied between the different data sets, yielding $N_{\mathrm{H}}=(1.98\pm0.07) \times 10^{22}~\nh$. Quoted uncertainties are at the 1$\sigma$ level of confidence.} 
\end{center}
\end{table*}


\subsection{Constraints on a Hard Emission Tail}\label{subsec:spec}
The quiescent spectra are soft and well-fitted with a thermal emission model. Previous analysis of the 2003/2009 and 2011 February/April data showed that any possible non-thermal emission tail, which is often seen in the spectra of quiescent neutron star LMXBs, could contribute at most $\simeq20\%$ to the total unabsorbed 0.5--10 keV flux \citep[][]{degenaar2011_terzan5_2,degenaar2011_terzan5,degenaar2011_terzan5_3}. We examined the possible presence of a hard spectral component by adding a power-law component (\textsc{pegpwrlw}) to the thermal model fits.

We first investigated the 2013 February data set, because it had the longest exposure time and the highest sensitivity. The \textsc{nsatmos} model provides an adequate fit ($\chi_{\nu}^2=1.25$ for 18 dof and $P_{\chi}$=0.21), but it can be seen in Figure~\ref{fig:spec} that the flux in the last energy bin is underestimated. Adding a power-law component results in $\chi_{\nu}^2=0.89$ for 16 dof and $P_{\chi}$=0.59. The photon index is not well constrained ($\Gamma<5.7$), but the best-fit value of $\Gamma=3.3$ is much softer than typically found for the quiescent spectra of neutron star LMXBs ($\Gamma \simeq1.5-2$). Given the limited quality of the spectral data, this is probably because the power law attempts to fit part of the thermal emission. This suggests that there is no significant power-law component present in the spectrum.

If we include a power-law component when fitting all data sets simultaneously, we again obtain a very soft index with large errors ($\Gamma=3.6^{+0.7}_{-1.3}$; yielding $\chi_{\nu}^2=0.67$ for 52 dof and $P_{\chi}$=0.97). To place limits on the power-law contribution we fixed the index to $\Gamma=2.0$. For this fit the fractional contribution of the hard tail to the unabsorbed 0.5--10 keV varies (non-monotonically) between 3\% and 22\% for the different data sets ($\chi_{\nu}^2=0.68$ for 53 dof and $P_{\chi}$=0.96). We consider these upper limits.


\subsection{The Crust Cooling Curve}\label{subsec:curve}
\begin{figure}
 \begin{center}
	\includegraphics[width=8.8cm]{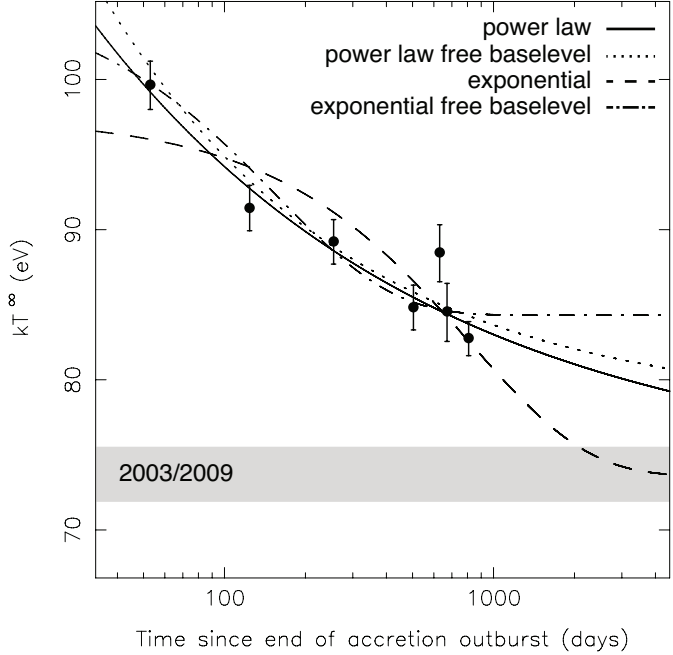}
    \end{center}
    \caption[]{Evolution of the neutron star temperature after the 2010 outburst along with decay fits. The solid and dashed lines represents fits to a power law and exponential decay that go down to the level detected in 2003/2009 (gray shaded area), respectively. The dotted (power law) and dash-dotted (exponential) lines are decay fits with the quiescent base level left as a free parameter. The end of the outburst was assumed to be 2010 December 26 \citep[MJD 55556;][]{degenaar2011_terzan5}. Error bars represent 1-$\sigma$ confidence intervals.
    }
 \label{fig:lc}
\end{figure}

Figure~\ref{fig:lc} displays the evolution of the temperature of the neutron star following the 2010 accretion outburst. The thermal emission follows a steady and smooth decay.\footnote[11]{The apparent enhancement in the 2012 September data (Obs ID 14475) is only at the $\simeq1\sigma$ level and hence not significant (see also Tables~\ref{tab:obs} and~\ref{tab:spec}).} This supports the hypothesis that the crust became significantly heated during the accretion phase and is cooling in quiescence. 

To characterize the shape of the crust cooling curve and allow for a comparison with other sources, we fit the temperature curve with an exponential decay function of the form 
$y(t)=a~e^{-(t-t_0)/\tau}+b$, and a power-law decay of the form $y(t)=a~(t-t_0)^{-\alpha}+b$. Here, $a$ is a normalization constant, $b$ a constant offset that represents the quiescent base level, $\tau$ the e-folding time, $\alpha$ the decay index, and $t_0$ the start time of the cooling curve \citep[assumed to be 2010 December 26, MJD 55556;][]{degenaar2011_terzan5}. 

First exploring the exponential decay, we find that fixing the quiescent base level to the temperature inferred from the 2003/2009 data (i.e., $b=73.6$~eV) results in a poor fit (dashed line in Figure~\ref{fig:lc}). If we instead allow this parameter to vary (i.e., assuming that the quiescent base level can differ between outbursts; see Section~\ref{subsec:alternative}), a better fit is obtained that yields a base level of $b=84.3\pm1.4$~eV (dash-dotted curve in Figure~\ref{fig:lc}; Table~\ref{tab:lc}). This is close to the value obtained from the 2013 February data ($kT^{\infty}=82.8\pm1.2$~eV) and suggests that the crust cooling curve may have (nearly) leveled off.

We obtain better fits by using a power-law decay function. Assuming a fixed base level of $b=73.6$~eV yields an acceptable fit (solid curve in Figure~\ref{fig:lc}; Table~\ref{tab:lc}). When including the base level as a fit parameter, we obtain $b=77.3\pm1.0$~eV. This is significantly lower than our most recent measurement of 2013 February (Table~\ref{tab:spec}). If the power-law fit is a correct description of the temperature evolution, it would therefore be indicative of continued cooling of the crust.

The crust cooling curves of other sources have been fit to a power-law decay without a constant offset \citep[][]{cackett2008,cackett2010,degenaar2010_exo2,fridriksson2011}. To allow for a direct comparison of the decay index of \igr, we therefore also report a fit without including a base level (i.e., assuming $b=0$; Table~\ref{tab:lc}).

\begin{table}
\begin{center}
\caption{Decay Fits to the Quiescent Light Curve\label{tab:lc}}
\begin{tabular*}{0.4\textwidth}{@{\extracolsep{\fill}}lc}
\hline\hline
Fit Parameter (unit) & Value  \\
\hline
\multicolumn{2}{c}{{\bf Exponential decay, base level fixed}}  \\
Normalization, $a$ (eV) & $23.8 \pm 1.3$   \\
Decay time, $\tau$ (days) & $825 \pm 107$  \\
Constant offset, $b$ (eV) & 73.6 fixed   \\
$\chi^2_{\nu}$ (dof) & 3.08 (5)   \\
$P_{\chi}$ & 0.01  \\
\hline
\multicolumn{2}{c}{{\bf Exponential decay, base level free}}  \\
Normalization, $a$ (eV)  & $21.6 \pm 4.0$   \\
Decay time, $\tau$ (days) & $157 \pm 62$  \\
Constant offset, $b$ (eV) & $84.3 \pm 1.4$   \\
$\chi^2_{\nu}$ (dof) & 1.84 (4)   \\
$P_{\chi}$ & 0.12  \\
\hline
\multicolumn{2}{c}{{\bf Power-law decay, base level fixed} } \\
Normalization, $a$ (eV) & $98.6 \pm 18.7$   \\
Decay index, $\alpha$ & $0.34 \pm 0.04$  \\
Constant offset, $b$ (eV) & 73.6 fixed   \\
$\chi^2_{\nu}$ (dof) & 1.21 (5)   \\
$P_{\chi}$ & 0.30   \\
\hline
\multicolumn{2}{c}{{\bf Power-law decay, base level free}}  \\
Normalization, $a$ (eV) & $147.9 \pm 12.7$  \\
Decay index, $\alpha$ & $0.47 \pm 0.05$  \\
Constant offset, $b$ (eV) & $77.3 \pm 1.0$   \\
$\chi^2_{\nu}$ (dof) & 1.20 (4)   \\
$P_{\chi}$ & 0.31   \\
\hline
\multicolumn{2}{c}{{\bf Power-law decay, no constant offset}}  \\
Normalization, $a$ (eV) & $124.8 \pm 4.8$  \\
Decay index, $\alpha$ & $0.06 \pm 0.01$  \\
Constant offset, $b$ (eV) & $0$ fix   \\
$\chi^2_{\nu}$ (dof) & 1.43 (5)   \\
$P_{\chi}$ & 0.21   \\
\hline
\end{tabular*}
\tablecomments{The quiescent data was fit to an exponential decay of the form $y(t)=a~e^{-(t-t_0)/\tau}+b$, and a power-law decay shaped as $y(t)=a~(t-t_0)^{-\alpha}+b$. The power-law fit without constant offset is included to allow for a direct comparison with other sources. The start of the cooling curve, $t_0$ was set to 2010 December 26 \citep[MJD 55556;][]{degenaar2011_terzan5}. Quoted uncertainties are at the 1$\sigma$ level of confidence.
}
\end{center}
\end{table}

\section{Discussion}\label{sec:discuss}
We use new \chan\ observations of the globular cluster Terzan 5 to further study the quiescent emission of the 11 Hz X-ray pulsar \igr. The new data cover a time span of $\simeq250-800$ days since the cessation of its 2010 October--December accretion phase. We combine these with two earlier observations obtained $\simeq$50 and $\simeq$125 days post-outburst. The source intensity is observed to decay smoothly over the 2.2~yr time span covered by the observations.

Fitting the spectral data with a neutron star atmosphere model suggests that the neutron star temperature steadily decreased by $\simeq20$\% from $kT^{\infty}\simeq100$ to $\simeq83$~eV. The inferred 0.5--10 keV luminosity decreased by a factor of $\simeq2.5$ from $L_{\rm X}\simeq1.5\times 10^{33}$ to $\simeq6.3\times 10^{32}~\lum$. The quiescent spectra of \igr\ are described well by a thermal model and there is no indication for the presence of a significant hard emission tail. By including a power-law spectral component to the thermal model fits, we found that it must always contribute $\lesssim22$\% to the total unabsorbed 0.5--10 keV flux.

The temperature determined from our most recent observations (2013 February) is higher than that measured in 2003/2009 at the $\simeq5\sigma$ level of confidence (Figure~\ref{fig:lc}). Likewise, the thermal flux remains a factor of $\simeq2$ above the pre-outburst level (Table~\ref{tab:spec}). If the source were to return to the 2003/2009 level, this suggests that the neutron star crust is still hot and needs to cool further.


\subsection{Comparison with Other Sources and Model Calculations}\label{subsec:compare}
The observed steady decrease in neutron star temperature provides strong support for the hypothesis that the crust was substantially heated during the 2010 accretion outburst and is currently cooling in quiescence. Such crustal cooling has previously been reported for four other neutron star LMXBs: \ks, \mxb, \exo, and \xte\ \citep[][]{wijnands2002,wijnands2004,cackett2008,cackett2010,degenaar09_exo1,degenaar2010_exo2,diaztrigo2011,fridriksson2010,fridriksson2011}. All four experienced prolonged outbursts lasting $\simeq1.5$ to $\simeq25$~yr, and show a continuous decrease in their quiescent thermal emission on a time scale of years. \igr\ is the first regular transient LMXB (i.e., with an outburst length of weeks to a few months) showing strong evidence for crustal cooling.\footnote[12]{Quiescent monitoring observations of another two sources recently commenced: \maxisource, which was active for $\simeq1.5$~yr in 2011--2012, and \terdrie\ in Terzan 5, which was active for $\simeq 2$~months in 2012. Data acquisition and analysis are in progress, but preliminary results reveal evidence for crustal cooling in both sources.}

It is instructive to compare the observed crust cooling curve with that of the quasi-persistent sources. Fitting the current data to a power-law decay without a constant offset results in a decay index of $\alpha=0.06\pm0.01$. Comparable values of $\alpha \simeq 0.03-0.07$ were obtained for \exo\ and \xte\ \citep[][]{degenaar2010_exo2,fridriksson2011}, whereas the decay indices of \ks\ and \mxb\ are higher \citep[$\alpha \simeq 0.13$ and $\simeq0.33$, respectively;][]{cackett2008,cackett2010}. If the crust cooling curves of the quasi-persistent sources are instead fitted with an exponential decay, this leads to e-folding time scales of $\tau \simeq200-500$~days \citep[][]{cackett2008,cackett2010,degenaar2010_exo2,fridriksson2011}. For \igr\ we obtain $\tau \simeq825$~days, assuming that it returns to its 2003/2009 level, although this decay fit does not match the data well. Nevertheless, we can conclude that despite that it was accreting for a significantly shorter time (by a factor $\gtrsim10$), the crust cooling curve is not strikingly different from that of the other four.

 \begin{figure}
 \begin{center}
	\includegraphics[width=8.8cm]{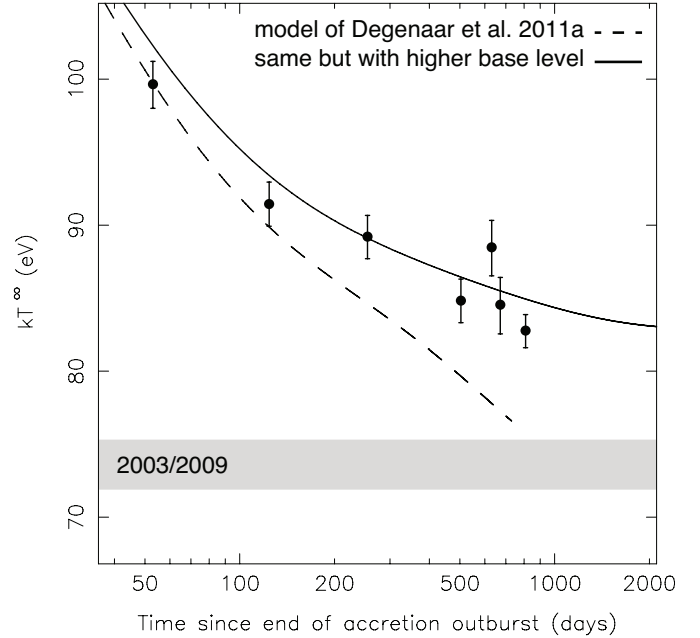}
    \end{center}
    \caption[]{Evolution of the neutron star temperature after the 2010 outburst compared to model calculations. The dashed line shows the model presented by \citet{degenaar2011_terzan5_3}, which was based on the first two data points. It included an extra crustal heat source of 1.0 MeV per accreted nucleon placed at a depth of $P/g=4.5\times10^{11}$~g~cm$^{-2}$ ($\rho \simeq 4\times10^{8}$~g~cm$^{-3}$) to match the temperature at early times. The solid curve shows the same model, but for a higher base temperature of $kT^{\infty}=84$~eV (compared to $kT^{\infty}=72$~eV for the dashed curve; see \citet{degenaar2011_terzan5_3} for details). 
    }
 \label{fig:model}
\end{figure}

It is remarkable that the crust cooling curve of \igr\ is rather similar to that of the quasi-persistent LMXBs and that the crust is still hot $\simeq2.2$~yr after the end of its outburst. Its much shorter outburst length of $\simeq11$ weeks (a factor of $\gtrsim10$ shorter than the others) should have caused significantly less heating, resulting in more rapid cooling \citep[e.g.,][]{page2012}. 
This is illustrated by Figure~\ref{fig:model}, where we compare the updated crust cooling curve with the model calculations presented in \citet{degenaar2011_terzan5_3}, which were based on the first two data points. 

In order to reach the observed high temperature at early times (within $\simeq125$~days after the outburst) an additional source of heat was needed at shallow depth in the crust \citep[at a column density of $P/g\lesssim10^{14}~\mathrm{g~cm}^{-2}$, corresponding to a matter density of $\rho \lesssim 3\times10^{10}~\mathrm{g~cm}^{-3}$;][]{degenaar2011_terzan5_3}. It is interesting to note that in an independent study, \citet{linares2012_ter5_2} found that the unusual thermonuclear X-ray burst activity of \igr\ might require additional energy release in the outer layers of the neutron star. The origin of such an additional heat source is currently unclear \citep[i.e., it is not accounted for by standard nuclear heating models; see the discussion in][]{degenaar2013_xtej1709}, but it was also invoked to explain the crust cooling curves of \ks\ and \mxb\ \citep[][]{brown08}. A recent observation of the neutron star LMXB \rx\ taken very shortly after an outburst also suggested the presence of a substantial heat source located in the outer crustal layers, although for that source any effects of possible ongoing accretion can not be excluded \citep[][]{degenaar2013_xtej1709}. 

It is clear from Figure~\ref{fig:model} that the preliminary calculations presented in \citet{degenaar2011_terzan5_3} do not match the new data points: the observed temperatures at later times are systematically higher than the model prediction. Possible ways to keep the neutron star hot for a longer time are to impose a lower thermal conductivity or a larger specific heat \citep[][]{rutledge2002,shternin07,brown08,page2012}. However, this could possibly make it more difficult to get the neutron star hot at early times, and would imply that the crust properties are different from that of the quasi-persistent neutron stars. It might therefore not be straightforward to explain our observations with standard heating and cooling models.


\subsection{Unusual Crust Properties?}\label{subsec:crustprop}
It is worth considering whether the neutron star in \igr\ may have unusual crust properties that can influence its thermal evolution. Apart from the short outburst, another feature that sets this source apart from the quasi-persistent LMXBs is that it showed X-ray pulsations (at 11 Hz) during outburst. This has three implications. Firstly, the neutron star likely has a higher magnetic field than the other four sources \citep[$B\simeq10^9-10^{10}$~G;][]{cavecchi2011,miller2011,papitto2010}. A high magnetic field can strongly affect the thermal evolution of neutron stars, although this is thought to become effective only at much higher field strengths of $B\gtrsim10^{12}$~G \citep[e.g.,][]{aguilera2008,pons2009,cooper2010}. Therefore, the magnetic field is not expected to be a source of influence in \igr.

Secondly, with a spin period of 11 Hz, the neutron star in \igr\ is rotating much slower than those in \ks, \mxb\ and \exo, which have spin periods of $\simeq$524, 567 and 552 Hz, respectively \cite[as inferred from the detection of oscillations during thermonuclear X-ray bursts;][the spin period of \xte\ is not known]{smith1997,wijnands01,galloway2010}. It was found by \citet{linares2012_ter5_2}, that the relatively low spin frequency of \igr\ has a profound effect on its thermonuclear X-ray bursting behavior (which are ignited in the surface layers of the neutron star). Rotation can also affect the structure of the crust and hence its heating and cooling properties \citep[e.g.,][]{haensel2008_rotation}. However, the effect of rotation on the crust equation of state is thought to become prominent only at high frequencies of $\gtrsim1000$~Hz and may therefore not be a source of influence for \igr\ and the other crust cooling neutron stars.

Thirdly, the fact that \igr\ harbors an 11 Hz pulsar indicates that the binary may have an unusual accretion history and started its Roche lobe overflow phase relatively recently \citep[$\lesssim10^{7}$~yr ago;][]{patruno2012}. It was noted by \citet{wijnands2012} that in such a young LMXB the neutron star might have a different crust composition. Accretion can replace the outer layers of the crust on a time scale of $\simeq10^{5}/\dot{M}_{9}$~yr, and the innermost crustal layers after $\simeq10^{7}/\dot{M}_{9}$~yr \citep[where $\dot{M}_{9}$ is the time-averaged mass-accretion rate of the binary in units of $10^{-9}~\mdot$;][]{chamel2008}. For an estimated long-term averaged accretion rate of $\simeq10^{-11}-10^{-10}~\mdot$ \citep[][]{wijnands2012}, the time needed to replace the crust in \igr\ would thus be $\simeq 10^{6}-10^{7}$~yr for the outer layers, and $\simeq 10^{8}-10^{9}$~yr for the inner ones. It is therefore conceivable that (part of) the crust is still composed of the original, catalyzed matter. This could markedly impact its thermal and transport properties.


\subsection{Alternative Explanations}\label{subsec:alternative}
There are alternative explanations that could possibly account for our observations. It is possible that matter continues to accrete onto the neutron star in quiescence. There is evidence for such low-level accretion in some neutron star LMXBs \citep[e.g.,][]{rutledge2002_aqlX1,campana2003_aqlx1,cackett2010_cenx4}. The resulting spectrum may be thermal and difficult to distinguish from that of a cooling neutron star \citep[][]{zampieri1995,soria2011}. Although little is understood about the physics of such a residual accretion flow, it is generally assumed that it would involve stochastic variability (on time scales of seconds to years) and the presence of a strong non-thermal emission component. Instead, the quiescent data of \igr\ show a very smooth decay and any possible hard spectral component can only contribute $\lesssim$22\% to the total unabsorbed 0.5--10 keV flux. There are therefore no obvious indications that low-level accretion is occurring.

As an accretion-heated crust cools, it eventually settles at a base level that is determined by the temperature of the core. Since the core temperature does not change appreciably in between different outbursts \citep[e.g.,][]{brown1998,colpi2001,ushomirsky2001}, the temperature of \igr\ would thus be expected to return to the level measured before the outburst in 2003/2009. However, it has been proposed that the heat flux flowing from the stellar interior to the surface is determined by the amount of hydrogen and helium that is left on the surface after the end of an outburst, and that this may change from one accretion phase to another \citep[][]{brown2002}. As a result, the observed thermal emission after different outbursts may differ by a factor of a few while the interior temperature is the same.

The quiescent light curve of \igr\ can be described by an exponential decay that levels off to a temperature of $kT^{\infty}=84.3\pm1.4$~eV. This is similar to the values determined from our last four observations (2012 May till 2013 February). 
This could suggest that the crust has already cooled, with a characteristic time scale of $\tau=157\pm62$~days. This is shorter than the decay times measured for the quasi-persistent sources and hence points to faster cooling, as would be expected for a shorter outburst length \citep[e.g.,][]{page2012}. Although a detailed theoretical modeling is beyond the scope of this paper, we briefly explored the effects of a higher base level on the crust cooling curve. The solid line in Figure~\ref{fig:model} shows the cooling trajectory using the same physics input as the dashed curve \citep[for details, see][]{degenaar2011_terzan5_3}, but with a higher quiescent base level of $kT^{\infty}=84$~eV. Such a model can better reproduce the shape of the observed cooling curve and could account for the higher temperatures at late times.\\

To conclude, it is possible that the quiescent emission of \igr\ has settled at a higher observed equilibrium level and that the neutron star crust has (nearly) cooled. In this case, the thermal flux and inferred neutron star temperature is not expected to change appreciably any more until a new outburst occurs (unless low-level accretion onto the neutron star occurs in quiescence, which would likely cause non-monotonic variability in the quiescent flux). However, the current data are better fit by a power-law decay, which is suggestive of continued cooling. If the temperature of the neutron star is indeed observed to decrease further, this could possibly challenge our current understanding of heating and cooling of transiently accreting neutron stars. Continued monitoring of Terzan 5 with \chan\ to further study the flux evolution of \igr\ can discriminate between the different possibilities. 

\acknowledgments
N.D. is supported by NASA through Hubble Postdoctoral Fellowship grant No. HST-HF-51287.01-A from the Space Telescope Science Institute. Support for this work was provided by NASA through Chandra Award Nos. GO2-13043X and G03-14034A issued by the Chandra X-ray Observatory Center. R.W. acknowledges support from a European Research Council starting grant. C.O.H. and G.R.S. are supported by Natural Sciences and Engineering Research Council Discovery Grants, and C.O.H. by an Ingenuity New Faculty Award.

{\it Facility:} \facility{CXO (ACIS)}

\end{document}